# Bismuth nanogratings with narrow plasmon resonances for dynamic polarized color generation and colorimetric sensing

Fernando Chacón-Sánchez[1], Fátima Cabello[1], Marina García Pardo[1], Emmanuel Haro-Poniatowski[2,1], Rosalía Serna[1] and Johann Toudert[1,*]

*1 Laser Processing Group, Instituto de Óptica, CSIC, Serrano 121, Madrid, Spain*

*2 Departamento de Física, Universidad Autonóma Metropolitana Iztapalapa, Apdo. Postal 55-534, Cd. De México, México*

*Corresponding author: johann.toudert@csic.es

**Abstract:** Bismuth nanostructures are appealing for sustainable color generation and sunlight harvesting, thanks to their non-toxicity and their tunable visible-to-near infrared interband plasmon resonances. However, owing to their broad and polarization-insensitive spectral features, the nanostructures reported so far displayed a limited color tunability and were lacking other functionalities such as sensing. Herein, we report bismuth nanogratings with polarization-sensitive and narrow plasmon resonances ($Q > 10$). They were fabricated on the $cm^2$ scale following a lithography-free approach: conformal pulsed laser deposition of bismuth onto both the reflective and transparent nanostructured layers of DVDs. We characterized their specular reflectance for different orientations of the plane of incidence, angles of incidence, and polarizations of light. When light is polarized in the plane perpendicular to the lines, plasmon resonances occur and shift across the visible-to-near infrared upon changing the angle of incidence. In contrast, no such resonances occur when light is polarized in the plane parallel to the lines. This results in well-contrasted, polarization-sensitive colors, which are iridescent for the former orientation of polarization, and not for the latter. Resonances strongly shift upon changing the refractive index of the surrounding medium (> 500 nm/RIU), resulting in a marked change in the plasmonic color, e.g., from green in air to red in water. This showcases the potential of these nanogratings for dynamic polarized color generation and colorimetric sensing.

**Keywords:** bismuth, plasmonics, nanogratings, polarization, color, sensing

Bismuth (Bi) nanostructures are appealing for sustainable color generation [1-5] and sunlight harvesting [5-6], thanks to their non-toxicity and their interband plasmon resonances [7-9], which are tunable by design across the ultraviolet, visible, and near-infrared [10-11]. However, the nanostructures reported so far displayed broad and polarization-insensitive spectral features, limiting their color tunability and making them unsuitable for other applications such as sensing. Therefore, to achieve enhanced functionalities and open the path to other Bi-based applications, it is needed to design Bi nanostructures with polarization-sensitive and narrow plasmon resonances. Furthermore, for their implementation on a practical scale, such nanostructures should be fabricated by cost-effective, lithography-free methods enabling the production of large amounts of material [12-14].

Narrow and polarization-sensitive plasmon resonances have been widely reported on metal nanogratings [15-19], which were initially known for their diffraction properties and Wood anomalies [20-21]. Metal nanogratings are nowadays fabricated in large amounts by cost-effective lithography-free methods such as nanoimprinting for applications in everyday life [22]. In particular, they are one of the key components of optical disks such as DVDs [23]. As shown in Figure 1a, DVDs consist of two main nanostructured layers stacked together: a metal (here, Ag) nanograting on a polycarbonate (PC) substrate, and a PC nanograting showing a complementary topography. By separating the two layers, one thus obtains a reflective metal nanograting and a transparent PC one. Hereafter, these nanostructured layers are named Ag/PC and PC, respectively. Because of their plasmonic properties, broad availability and low cost, the so-obtained metal nanogratings have been repurposed for applications beyond data storage, such as light harvesting or sensing [24-30].

In this work, we have repurposed the two mentioned nanostructured layers of DVDs as templates for the conformal pulsed laser deposition of Bi, in order to obtain optically-thick Bi nanogratings on two different substrates, as shown in Figure 1a. The followed approach, which is described in the Methods section, is fully lithography-free and enables fabrication on the cm$^2$ scale in the laboratory. The so-obtained nanogratings are named Bi/Ag/PC and Bi/PC. They display narrow and polarization-sensitive plasmon resonances (with a quality factor Q > 10) in the visible-to-near infrared. This behavior is enabled by the negative value of the permittivity of Bi ($\varepsilon_1$ < 0), which takes values comparable with that of Ag in this spectral region as shown in Figure 1b, and by the nanograting dimensions (period ~ 700 nm, height ~ 70 nm).

The optically resonant character of the Bi/Ag/PC and Bi/PC nanogratings is demonstrated by the FDTD-calculated reflectance spectra at normal incidence shown in Figure 1c, which display a narrow resonance (fwhm ~ 50 nm, Q ~ 15) at the wavelength of 745 nm, when the incident light is polarized in the plane perpendicular to the lines ($E_0\perp$). The plasmonic nature of this resonance is demonstrated by the corresponding FDTD field maps of the Bi/Ag/PC nanograting shown in Figure 1d, where a resonant near-field enhancement occurs for the $E_z$ electric component at the wavelength of 745 nm. This plasmonic character of the resonance is further demonstrated by its important FDTD-calculated wavelength shift as a function of the refractive index of the surrounding medium, with a sensitivity higher than 700 nm/RIU (Supporting Information S1). Polarization-sensitivity is demonstrated in Figure 1c, which shows that no resonance occurs when light is polarized in the plane parallel to the lines ($E_0//$), since plasmons associated to the grating are unable to couple with a transverse electric field.

Note that the Bi/Ag/PC and Bi/PC nanogratings display resonances with slightly different widths and intensity. This is because we assumed different topographies for these two gratings (see Methods section), in accordance with the topographies of the actual fabricated nanogratings described hereafter. Note also that the observed resonance for the Bi/Ag/PC nanograting is broader and more intense than that of the Ag/PC substrate with the same topography, because of the higher optical losses ($\varepsilon_2$) of Bi compared with Ag (Figure 1b). In addition, no resonance is observed for the PC substrate, because its optically dielectric properties ($\varepsilon_1 > 0$, Figure 1b) do not enable it to support plasmons.

To experimentally validate the FDTD-calculated features, we characterized the structure and the optical properties of the fabricated nanogratings. The scanning electron microscope images along with the atomic force microscopy profiles displayed in Figure 2a show that the two nanogratings have the same ~ 700 nm period but a different topography: elliptical and rounded for the Bi/Ag/PC nanograting and trapezoidal with sharper edges for the Bi/PC one, yet with comparable heights: 80 and 70 nm, respectively. These features result from the conformal growth of Bi on the underlying substrates, thus replicating their period and topography. This conformal growth is better seen in the tilted view scanning electron microscopy images shown in the Supporting Information S2. These images also reveal the nanoscale roughness of the gratings, which likely results from the polycrystalline nature of the deposited material.

Both nanogratings are optically opaque in the visible-to-near infrared. We characterized their specular reflectance in the visible-to-near infrared, using a spectroscopic ellipsometer. Measurements were done with different orientations of the plane of incidence, angles of incidence, and polarizations of light. Unless specified, optical measurements were done in air, in ambient conditions. Figure 2b displays the obtained reflectance spectra at a 20º angle of incidence (the smallest achievable angle with this measurement setup) for the incident light being polarized in the plane perpendicular to the lines ($E_0\perp$, to ensure that plasmon resonances are observed). The spectra correspond to two different orientations of the plane of incidence (that contains the wavevector or the incident light $k_0$): parallel to the lines ($k_0//$) or perpendicular to them ($k_0\perp$), as depicted on the diagram in Figure 2b. For both the Bi/Ag/PC and Bi/PC nanogratings, distinct narrow resonances are observed in the $k_0//$ and $k_0\perp$ orientations: one resonance ($1^{//}$) at the wavelength of 700 nm for the $k_0//$ orientation, and two resonances ($1^\perp$ and $2^\perp$) at the wavelengths of 520 and 1000 nm for the $k_0\perp$ orientation. These multiple resonances appear at both longer and shorter wavelengths than the resonance observed at 745 nm in the simulations done at normal incidence (Figure 1c), because of the well-documented splitting of the nanograting resonances at oblique incidence [15]. Moreover, as predicted by the simulations, the resonances of the Bi/Ag/PC nanograting are broader and more intense than those of the Ag/PC substrate, and no resonances are observed for the PC substrate.

Interestingly and surprisingly, while the resonances of the Bi/Ag/PC nanograting appear as minima in the spectrum as in Figure 1c, the Bi/PC nanograting shows maxima at the corresponding wavelengths, over a broadband background with a much lower reflectance. Such background is particularly low for the $k_0\perp$ orientation. We attribute this mirrored behavior between the two nanogratings to a distinct interaction of light with them, which likely results from their complementary topography [16,17] and the different orientation of the bismuth crystal lattice [33].

To further characterize this interaction, we measured the diffuse reflectance spectra of the nanogratings in the visible range using a spectrophotometer equipped with an integrating sphere (at normal incidence, imposed by the setup configuration). These measurements yield diffuse reflectance values between 20% and 60%, showing that the nanogratings significantly reflect light in non-specular directions together with reflecting it specularly (Supporting Information S3). Importantly, the diffuse reflectance values are different for the Bi/Ag/PC and Bi/PC nanogratings. This is particularly notorious for the $k_0\perp$ orientation, for which diffuse reflectance is markedly higher for the Bi/PC nanograting than for the Bi/Ag/PC one, across a broad spectral region.

This leads us to tentatively propose that the mirrored behavior observed in the specular reflectance spectra results from a different trade-off between optical absorption, specular reflection, and non-specular reflection for the two nanogratings. For the Bi/Ag/PC nanograting, specular and non-specular reflection have a comparable importance across the visible spectrum, and having a marked absorption on the resonances reduces the specularly reflected signal, leading to a minimum in the spectrum. In contrast, for the Bi/PC nanograting, non-specular reflection overcomes the specular one across the spectrum, although the latter is enhanced on resonance leading to a maximum. Such mirrored behavior was not seen with the FDTD simulations (Figure 1c), because the spectra were computed by integrating both the diffuse and specular contributions indistinctively, and at normal incidence where the optical contrast between the two gratings is likely less important.

To fully address the sensitivity of the plasmon resonances on the angle of incidence (AOI), polarization ($E_0\perp$ or $E_0//$) and orientation of the plane incidence ($k_0//$ or $k_0\perp$), we report the specular reflectance spectra of the fabricated nanogratings measured as a function of these three parameters altogether. The corresponding maps showing the reflectance of the Bi/Ag/PC and Bi/PC nanogratings as a function of the wavelength and AOI are gathered in Figure 3. The maps of the reference Ag/PC and PC substrates are shown in the Supporting Information S4 for comparison. Figure 3 confirms that, for any AOI, the resonances of both the Bi/Ag/PC and Bi/PC nanogratings occur only for the $E_0\perp$ polarization. They further split upon increasing the AOI: the $1^{//}$, $1^{\perp}$ and $2^{\perp}$ resonances blue-shift from 700 to 400 nm, red-shift from 1000 to 1400 nm and blue-shift from 520 to below 400 nm, respectively. An additional, second-order resonance ($3^{\perp}$) appears at higher AOIs and red-shifts from 550 to 900 nm. All the resonances observed for the Bi/PC nanograting mirror those of the Bi/Ag/PC one, i.e., they lead to maxima in the spectrum, while those of Bi/Ag/PC appear as minima.

Such strong sensitivity of the plasmon resonances markedly impacts the perceived color of the nanogratings. To evaluate this impact, we calculated the CIE-1931 coordinates for the reflectance of both the Bi/Ag/PC and Bi/PC nanogratings as a function the AOI, polarization ($E_0\perp$ or $E_0//$) and orientation of the plane incidence ($k_0//$ or $k_0\perp$). All calculations were done assuming that the nanogratings are shined and observed in specular reflectance conditions. The specular reflectance spectra used as input for the calculations are shown in the Supporting Information S5. The color trajectory in the chromaticity diagram as function of the AOI is displayed in Figure 4a, and the corresponding colors are represented in Figure 4b. The polarization sensitivity of the plasmon resonances yields well-contrasted polarization-sensitive colors. The occurrence of such resonances for the $E_0\perp$ polarization and their shift with the AOI result in iridescent colors, while their absence for the $E_0//$ polarization results in an almost non-iridescent, orange color. Iridescence is much clearer in the case of the Bi/PC nanograting. For this nanograting, a wide gamut of vivid colors from red to violet is achieved upon varying the AOI between 20º to 70º. These vivid colors are enabled because the plasmon resonances of the Bi/PC nanograting yield reflectance maxima over a low broadband background, as discussed in the previous paragraphs.

Based on the knowledge of the plasmonic response and color appearance of the Bi/Ag/PC and Bi/PC nanogratings, we performed a simple experiment to demonstrate their potential for colorimetric sensing and dynamic color tuning. The change in their specular reflectance spectrum when immersing them in water was measured for an AOI of 45º with $E_0\perp$ polarized light and a $k_0//$ orientation of the plane of incidence. The sample was installed in a hollow prism that was either empty or filled with distilled water, as shown in Figure 5a. In these conditions, the reflectance spectrum of the nanograting (either Bi/Ag/PC or Bi/PC) in air is dominated by the $1^{//}$ resonance at visible wavelengths. Upon immersing the nanograting in water, its resonance red-shifts of about 170 nm, resulting in a high sensitivity of 515 nm/RIU, as shown in Figure 5b. This important shift results in a change in specular color, which is displayed in Figure 5c: from light pink to light orange for the Bi/Ag/PC nanograting, and from green to red for the Bi/PC nanograting. More vivid colors and a stronger color change are observed in the case of the Bi/PC nanograting, because its plasmon resonance results in a reflectance maximum over a low background, while that of the Bi/Ag/PC nanograting results in a reflectance minimum.

Summarizing, we have demonstrated that Bi nanogratings deposited by pulsed laser deposition onto the two different layers of DVD templates display narrow and polarization-sensitive plasmon resonances in the visible-to-near-infrared. Such resonances occur when the electric field of the incident light is polarized in the plane perpendicular to the lines, and not when polarized in the plane parallel to them. They display strong spectral shifts as a function of the angle of incidence and their wavelength also depends on the orientation of the plane of incidence. As a consequence, the nanogratings display polarization-sensitive colors in specular reflectance conditions, with iridescent (resp. almost non-iridescent) colors when polarization is oriented in the plane perpendicular (resp. parallel) to the lines. The strongest polarization-induced color contrast, and the most vivid and iridescent colors are observed for the Bi nanograting grown on the PC layer, because of its particular spectral response: its resonances result in reflectance maxima (instead of reflectance minima for the Bi nanograting grown on the Ag/PC layer). The resonances are strongly sensitive to the refractive index of the surrounding medium, thus enabling a clearly naked-eye visible change in the nanograting color only by immersing it in water.

Therefore, this study suggests the potential of Bi nanogratings for colorimetric sensing, where changes in the gratings' environment result in a sizeable color change. In this context, the gratings grown on the PC layer appear the most suitable. Furthermore, because of the purity of the obtained colors and their sensitivity to polarization, the nanogratings appear also suitable for dynamic polarized color generation. Towards their integration in a device, the generated color could be tuned or switched either by rotating the polarization of the incident light, changing the angle of incidence, rotating the nanograting in its plane, or modulating the refractive index of the surrounding medium. This versatility provides a more efficient and robust tuning mechanism than previous approaches that relied on tailoring the material design [1,2,11] or triggering temperature or light-induced changes in the dielectric function of Bi [34-37]. Implementing and testing the nanogratings for such applications will be facilitated by the simple and lithography-free approach followed to fabricate them. It enabled the production of samples at the cm$^2$ scale in our laboratory, and can be upscaled for larger area fabrication using broadly available and cheap DVD-R as templates. In this context, Bi will be more advantageous than other metals because of its non-toxicity – it is known as the "green metal" by the catalysis community – and low cost compared with noble metals (at present, Bi: 200 euros/kg, Ag: 1600 euros/kg, Au: 100000 euros/kg).

**Author contribution**

**F.C.S.** nanograting fabrication, structural characterization, discussion. **F.C.** optical characterization, discussion. **M.G.P.** sensing, discussion. **E.H.-P.** nanograting fabrication, sensing, discussion. **R.S.** funding acquisition, structural characterization, discussion. **J.T.** concept proposal, organization of experiments, writing initial draft, preparing graphical content, optical modeling, optical characterization, colorimetry, sensing, discussion.

**Acknowledgements**

This work has been partly funded by the national research grants ALPHOMENA (PID2021-123190OB-I00f) and SLIM-2P (PID2024-156974OB-C21) funded by MCIN/AEI/10.13039/501100011033 and by the European Union NextGenerationEU/PRTR and the European Regional Development Fund (ERDF).

**Methods**

*FDTD simulations.* Stable and converged 2D simulations were done with the OPTIFDTD software. The unit cell consisted of one period of the nanograting in the horizontal direction x and its height was chosen so that it includes the whole near-field and transition region of the nanograting. Periodic boundary conditions and perfect matching layers were chosen in the horizontal direction x and vertical direction z, respectively. The incident light was sent from a plane located above the grating, at normal incidence. The field maps were calculated using a monochromatic sinusoidal incident wave. To calculate the spectra, a Gaussian modulated light pulse was sent onto the nanograting, and the reflected light was collected in a plane located at the top of the unit cell, in the far-field region of the nanograting. The spectra were then determined from the Fourier transform of the collected signal. With such configuration, the calculated spectra include indistinctively the contributions of specularly reflected light, diffuse scattered, and diffracted light. The Bi/Ag/PC and Bi/PC nanogratings had a 700 nm period, a 70 nm height, and the Bi layer was 70 nm thick. The Bi/Ag/PC nanograting was modeled as a Bi/Ag stack, in which both layers had conformal topographies with an elliptic cross-section. They are optically thick, so that the underlying PC substrate could be omitted. The ellipse had a 70 nm half-length along its short axis (representing the grating height) and its long axis length was 640 nm, so that the distance between ellipses in the periodic structure was 120 nm. The Bi/PC nanograting was modeled as a Bi/PC stack in which both layers have conformal topographies with a trapezoidal cross-section. The top and bottom lengths of the trapeze were 200 nm and 500 nm respectively, so that the distance between trapezes in the periodic structure was 200 nm. The same topographies and dimensions were used for the Ag/PC and PC substrates. The dielectric functions of Bi, Ag and PC were taken from [9], [31], [32], respectively.

*Nanograting fabrication and optical characterization.* The Ag/PC and PC substrates were prepared by separating the two layers of commercial DVD-R disks. Sections of the disks were cut with scissors. Using a cutter, each section was separated into the two layers, which were cleaned in an ethanol bath and dried with nitrogen to remove the violet lacquer layer. The substrates had dimensions of few cm$^2$. They were used as template for the conformal pulsed laser deposition of Bi at room temperature in vacuum. The deposition setup and deposition conditions were the same as in our previous works. The nominal thickness of deposited Bi was 45 nm and 65 nm for the Bi/Ag/PC and Bi/PC nanogratings, respectively, as determined by spectroscopic ellipsometry on control thin films grown onto flat Si substrates during the same runs. Films with such thicknesses are optically-thick. Specular reflectance was measured with a Woollam VASE ellipsometer. Diffuse reflectance was measured with a Varian Cary 5000 spectrophotometer equipped with an integrating sphere and a polarizer. AFM profiles were obtained with a Park XE7 microscope with OML-AC160TS cantilevers in non-contact mode.

**Figures**

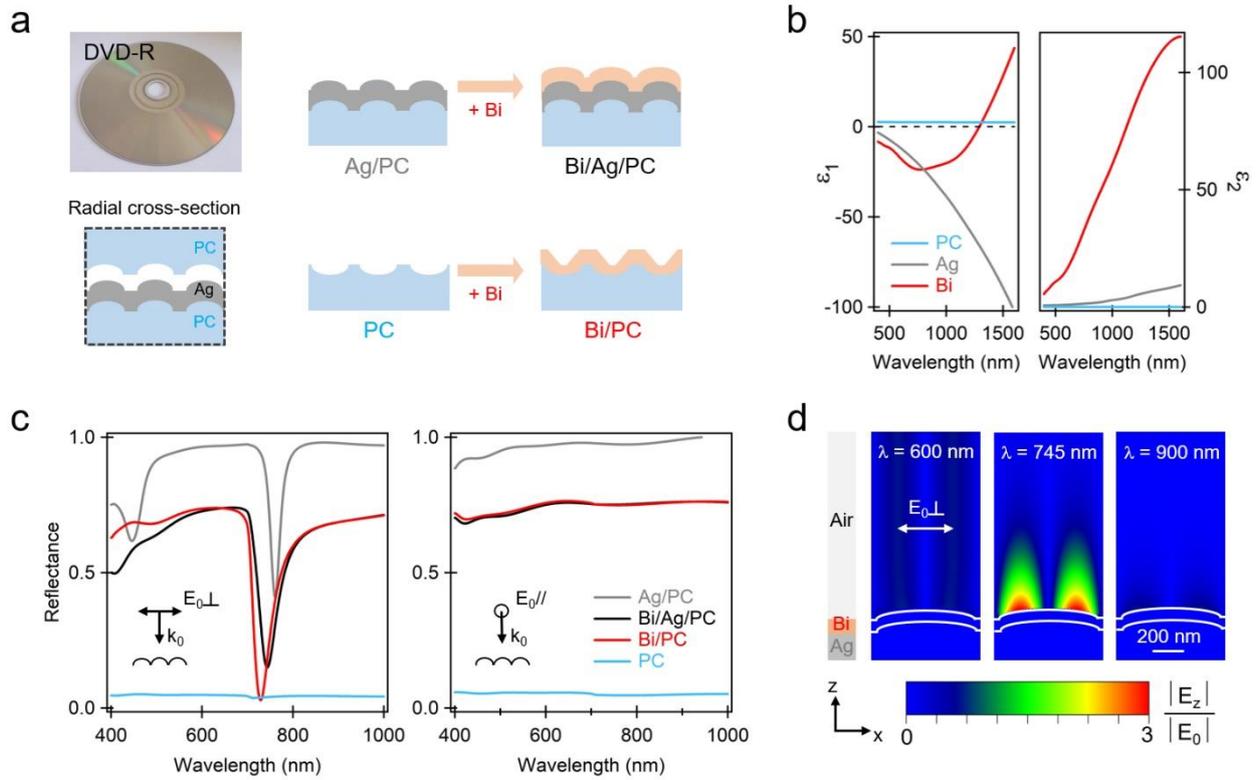

**Figure 1. Bismuth nanogratings deposited on DVD templates: schematic structure and calculated narrow and polarization-sensitive plasmon resonance.** (a) Picture of a DVD-R. Drawing of its cross-section perpendicular to the radius, showing the two main nanostructured layers: reflective Ag/PC and transparent PC. Process followed to deposit the Bi nanogratings: the two layers are mechanically separated and Bi is grown conformally onto them by pulsed laser deposition. By such means, two different nanogratings are obtained: Bi/Ag/PC and Bi/PC. (b) Dielectric functions of Bi, Ag, and PC, taken from [9], [31], [32], respectively. (c) FDTD-calculated reflectance spectra (in air) of the two nanogratings (period 700 nm, height 70 nm, Bi thickness 70 nm) and the corresponding Ag/PC and PC substrates, at normal incidence for the electric field of the incident light polarized in the plane perpendicular ($E_0\perp$) or parallel ($E_0//$) to the lines. A narrow plasmon resonance occurs in Bi/Ag/PC, Bi/PC and Ag/PC only for the $E_0\perp$ polarization. (d) FDTD-calculated maps of the vertical component of the electric field ($E_z$) of the Bi/Ag/PC nanograting at different wavelengths, showing a plasmonic field pattern at the wavelength of 745 nm.

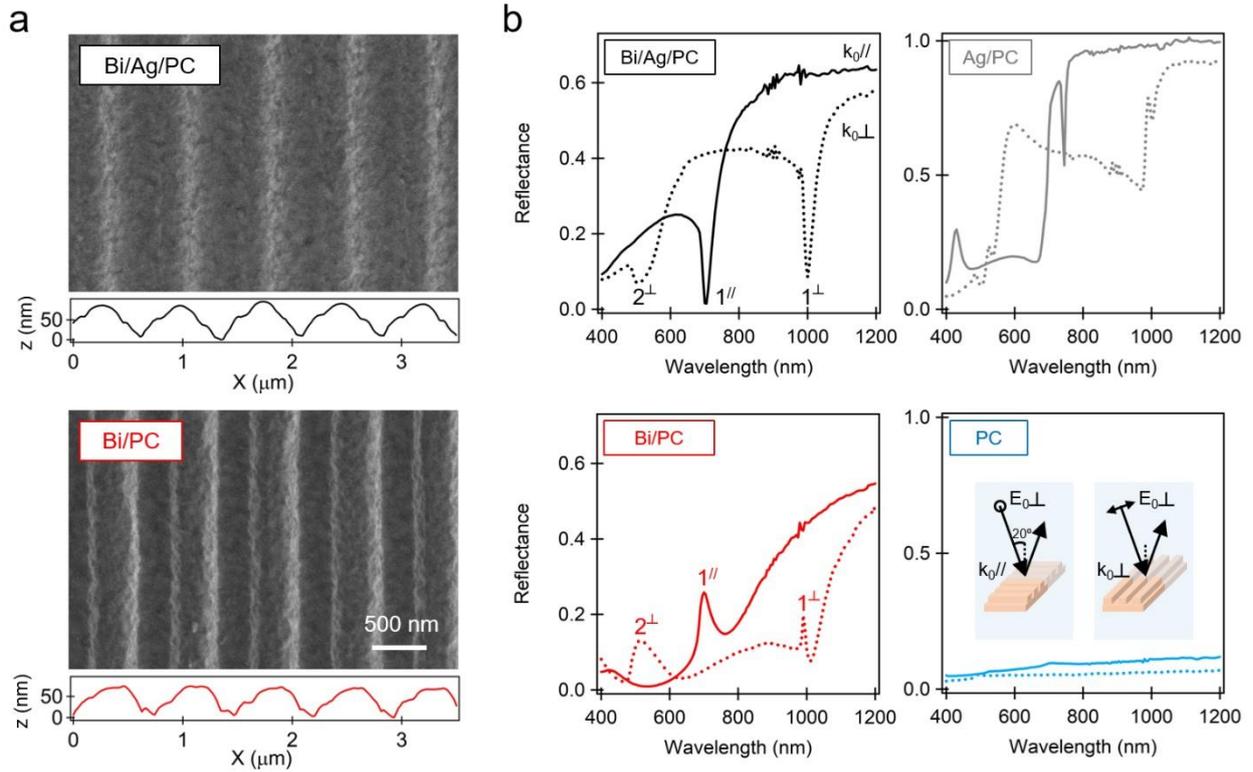

**Figure 2. Structure and narrow plasmon resonances of the fabricated nanogratings.** (a) Top-view scanning electron microscope images and atomic force microscopy profiles of the fabricated Bi/Ag/PC and Bi/PC nanogratings, evidencing their different topography. (b) Experimentally measured specular optical reflectance spectra (in air) of the two nanogratings and of the Ag/PC and PC substrates, at an angle of incidence of 20º, for the electric field of the incident light polarized in the plane perpendicular ($E_0\perp$) to the nanograting, and for the plane of incidence oriented parallel to the lines ($k_0/\!/$) or perpendicular to them ($k_0\perp$). Several narrow resonances ($1^{/\!/}$, $1^\perp$ and $2^\perp$) are evidenced for Bi/Ag/PC, Bi/PC and Ag/PC. Different resonances are seen for the different orientations of the plane of incidence. They result from the splitting of the normal-incidence resonance. The resonances are mirrored for the Bi/Ag/PC and Bi/PC nanogratings, for which they appear as minima and maxima in the spectra, respectively.

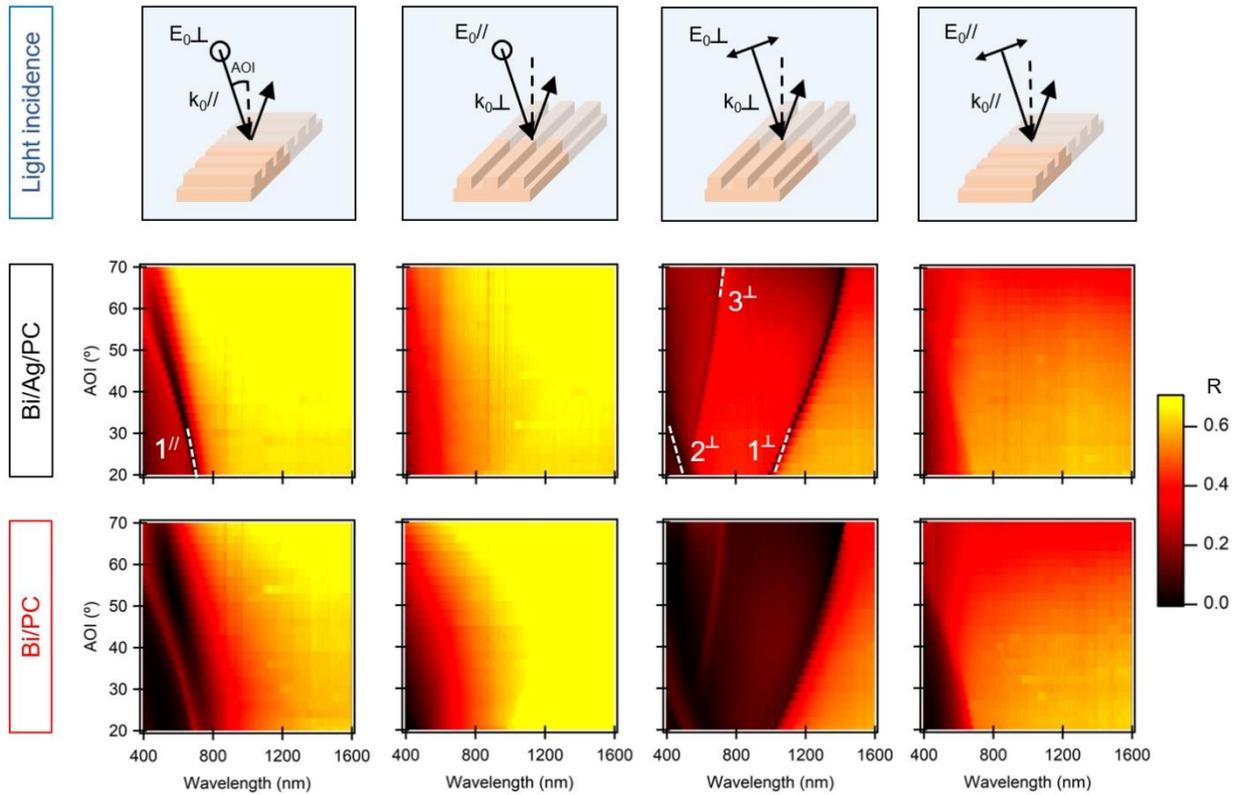

**Figure 3. Sensitivity of the narrow resonances of the fabricated nanogratings on the angle of incidence, for different polarizations and orientations of the plane of incidence.** The light incidence conditions are represented in the upper line (polarization: $E_0\perp$ or $E_0/\!/$, orientation of the plane of incidence: $k_0/\!/$ or $k_0\perp$). The two following lines represent color maps of the specular reflectance R (in air) of the Bi/Ag/PC and Bi/PC nanogratings as a function of wavelength and angle of incidence (AOI), for the light incidence conditions depicted on the first line. The resonances further split when the angle of incidence is increased, and shift across the whole visible to near-infrared. They are mirrored for the Bi/Ag/PC and Bi/PC nanogratings.

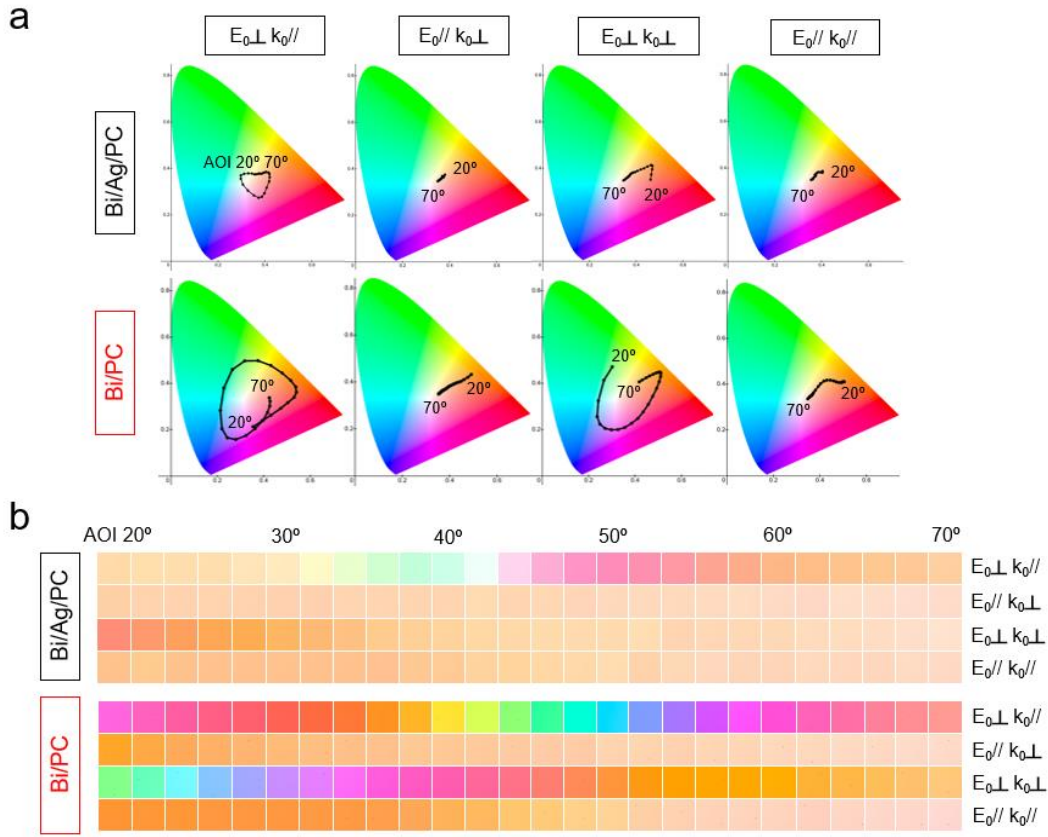

**Figure 4. Color of the fabricated nanogratings: polarization-sensitivity.** (a) CIE-1931 coordinates of the Bi/Ag/PC and Bi/PC nanogratings in specular reflectance conditions (in air), as a function the angle of incidence (AOI), polarization ($E_0\perp$ or $E_0/\!/$) and orientation of the plane incidence ($k_0/\!/$ or $k_0\perp$). (b) Corresponding colors of the Bi/Ag/PC and Bi/PC nanogratings, as a function of the AOI. Clear color contrasts are seen when changing polarization from $E_0\perp$ or $E_0/\!/$, because a plasmon resonance occurs in the former case and not in the latter. This results in iridescent colors - from red to violet - in the former case, and an almost non-iridescent orange color in the latter. More vivid colors and a stronger polarization-sensitivity are seen for the Bi/PC nanograting because its plasmon resonances yield maxima in the reflectance spectra instead of minima.

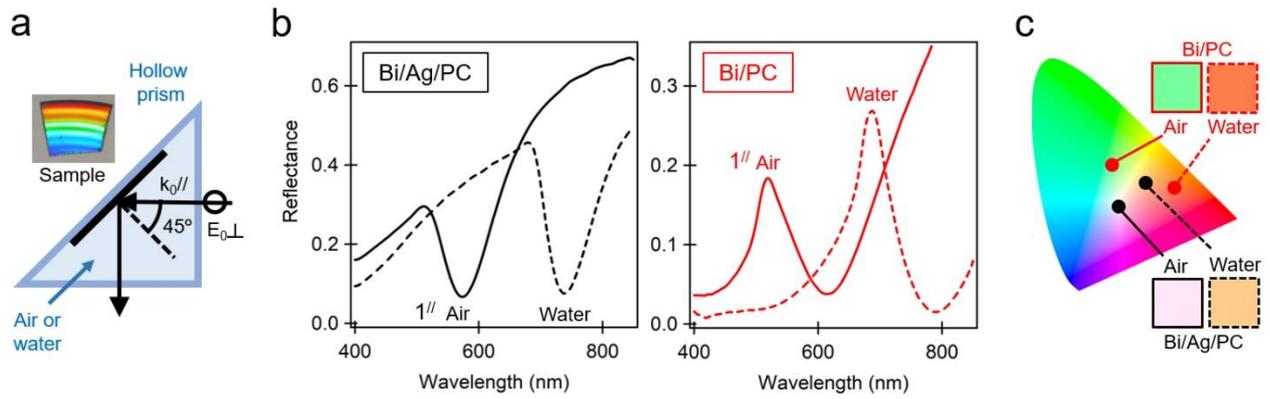

**Figure 5. Application of the fabricated nanogratings for colorimetric sensing and dynamic polarized color generation.** (a) The nanograting (Bi/Ag/PC or Bi/PC) is placed in a hollow prism. $E_0\perp$ polarized light is shined at an AOI of 45º, with a $k_0//$ orientation of the plane of incidence. In these conditions, the $1^{//}$ plasmon resonance occurs at visible wavelengths when the prism is filled with air. (b) Experimentally measured specular reflectance spectra for both nanogratings in air, and then immersed in water, showing a marked 170 nm red-shift (sensitivity: 515 nm/RIU). (c) Corresponding specular color change: from light pink to light orange for Bi/Ag/PC, and from green to orange for Bi/PC. More vivid colors and a stronger color change are seen for the Bi/PC nanograting because its plasmon resonance yields a maximum over a low background in the reflectance spectrum, while that of the Bi/Ag/PC nanograting appears as a minimum.

# Supporting Information

## S1. Refractive index sensitivity of the plasmon resonance

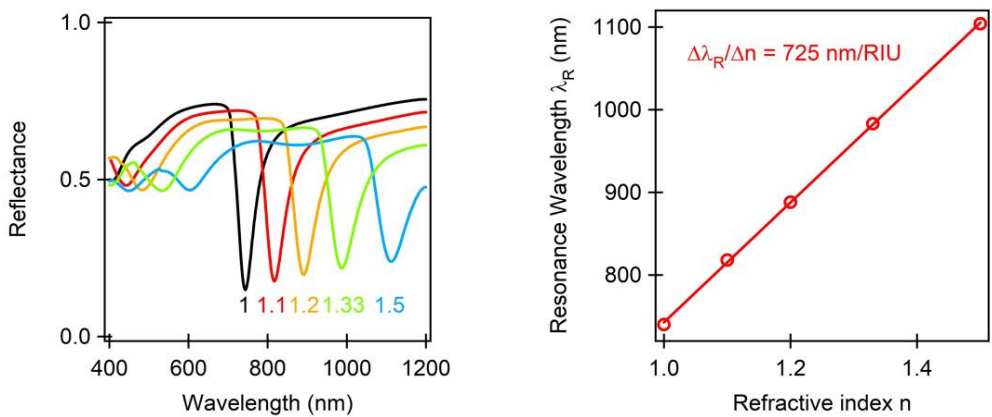

**Figure S1. FDTD-calculated sensitivity of the plasmon resonance in the Bi/Ag/PC nanograting.** Reflectance spectra (in air) of the nanograting at normal incidence for the $E_0\perp$ polarization, for different values (n from 1 to 1.5) of the refractive index of the surrounding medium. Corresponding shift of the plasmon resonance wavelength $\lambda_R$ as a function of n, yielding a theoretical sensitivity of 725 nm/RIU.

## S2. Additional scanning electron microscopy images

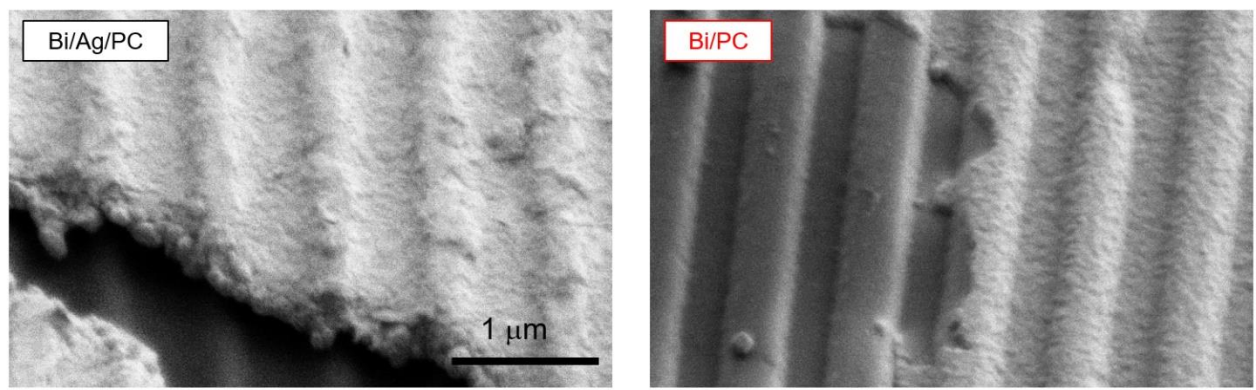

**Figure S2. Tilted-view scanning electron microscopy images of the Bi/Ag/PC and Bi/PC nanogratings.** The images have been taken at the vicinity of a scratch, in order to image both the Bi nanograting and the underlying substrate. For the Bi/PC nanograting, the PC substrate appears in dark contrast, and the Bi covered region in clear contrast. For the Bi/Ag/PC nanograting, the area in clear contrast corresponds to the Bi/Ag bilayer, while the area in dark contrast correspond to the PC. These images illustrate the conformal Bi growth and the different topographies of the two nanogratings.

## S3. Diffuse reflectance

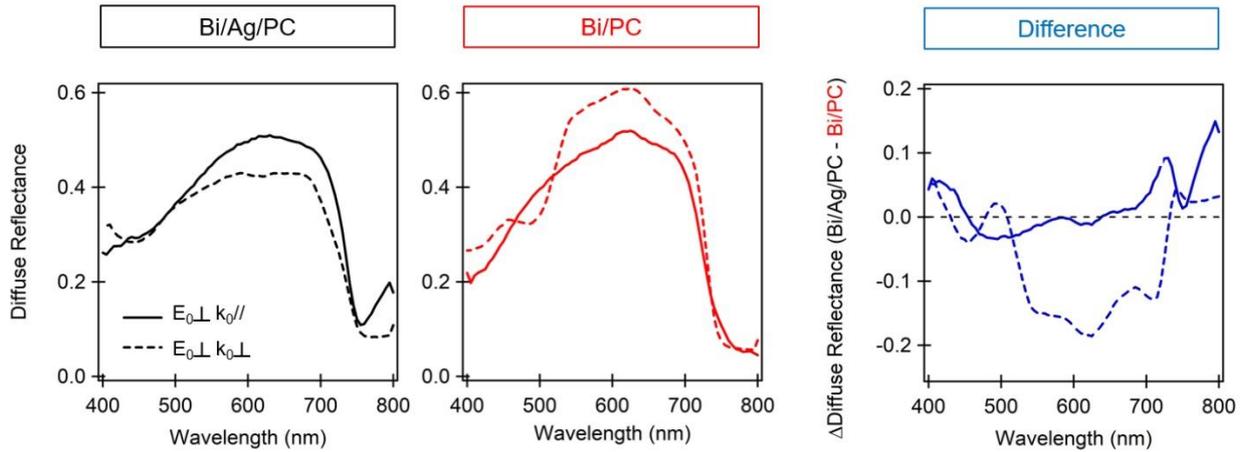

**Figure S3. Diffuse reflectance of the Bi/Ag/PC and Bi/PC nanogratings.** Measurements were done in air using an integrating sphere, at normal incidence with $E_0\perp$ polarized light, for two different orientations of the plane of incidence, $k_0//$ and $k_0\perp$. The spectra obtained for the two nanogratings are shown together with the difference between them.

## S4. Optical properties of the substrates

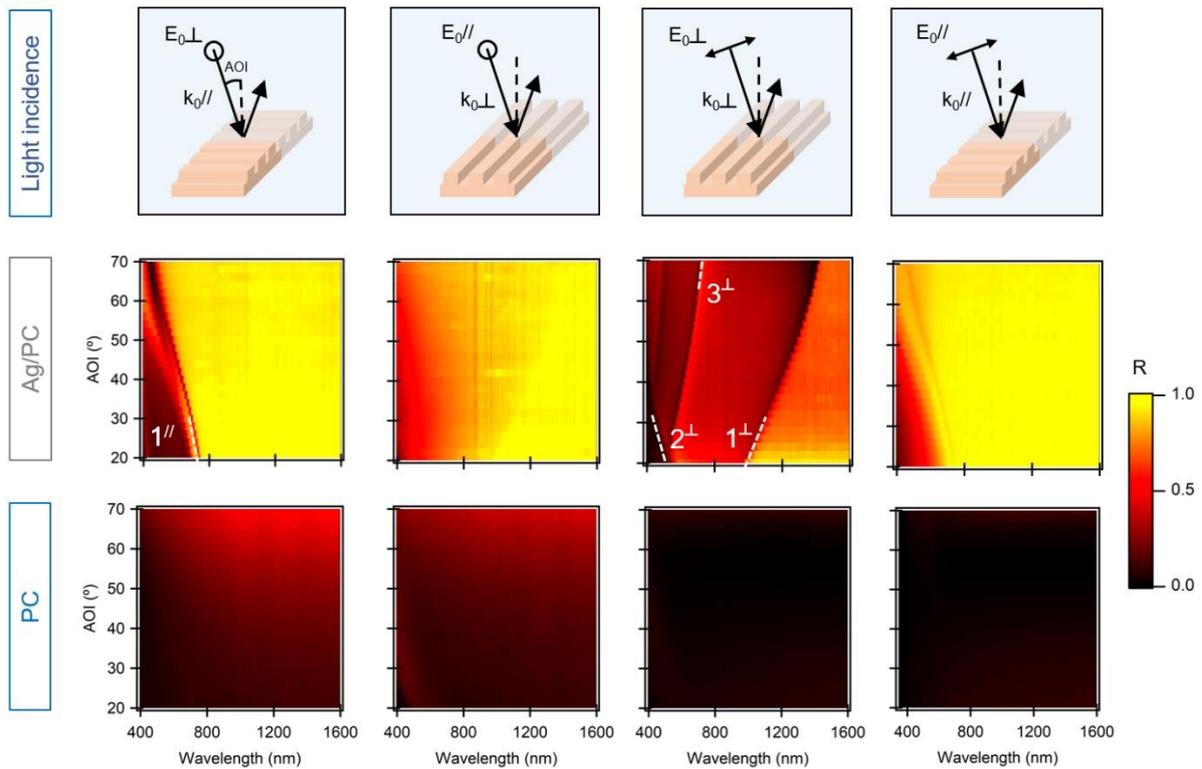

**Figure S4. Specular reflectance of the Ag/PC and PC substrates.** The light incidence conditions are represented in the upper line (polarization: $E_0\perp$ or $E_0//$, orientation of the plane of incidence: $k_0//$ or $k_0\perp$). The two following lines represent color maps of the reflectance R (in air) of the two substrates as a function of wavelength and angle of incidence (AOI), for the light incidence conditions depicted on the first line.

## S5. Optical data used in the CIE-1931 and color calculations

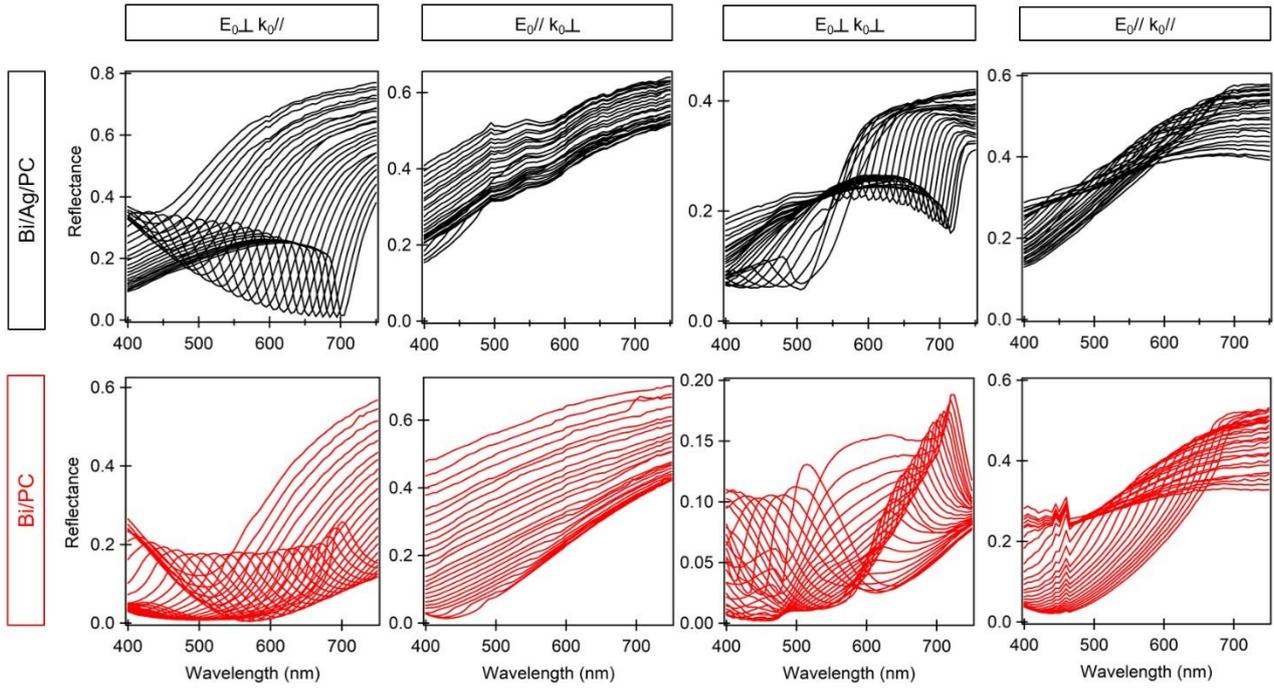

**Figure S5. Specular reflectance spectra of the Bi/Ag/PC and Bi/PC nanogratings.** Each graph represents the spectra as a function of the AOI. These spectra were measured in air. The different graphs correspond to different sets of polarization ($E_0\perp$ or $E_0//$) and orientation of the plane of incidence ($k_0//$ or $k_0\perp$).